\documentclass[twocolumn,preprintnumbers,showpacs,amsmath,amssymb]{revtex4}
\usepackage{graphicx}

\begin{document}
\title{Can relativistic bit commitment lead to secure quantum oblivious
transfer?}
\author{Guang Ping He}
\email{hegp@mail.sysu.edu.cn} \affiliation{School of Physics and
Engineering, Sun Yat-sen University, Guangzhou 510275, P. R. China}

\begin{abstract}
While unconditionally secure bit commitment (BC) is considered impossible
within the quantum framework, it can be obtained under relativistic or
experimental constraints. Here we study whether such BC can lead to secure
quantum oblivious transfer (QOT). The answer is not completely negative. On
one hand, we provide a detailed cheating strategy, showing that the
\textquotedblleft honest-but-curious adversaries\textquotedblright\ in some
of the existing no-go proofs on QOT still apply even if secure BC is used,
enabling the receiver to increase the average reliability of the decoded
value of the transferred bit. On the other hand, it is also found that some
other no-go proofs claiming that a dishonest receiver can always decode all
transferred bits simultaneously with reliability $100\%$ become invalid in
this scenario, because their models of cryptographic protocols are too ideal
to cover such a BC-based QOT.
\end{abstract}

 \maketitle



\section{Introduction}

Besides the well-known quantum key distribution (QKD) \cite{qi365}, bit
commitment (BC) and oblivious transfer (OT) are also essential cryptographic
primitives. It was shown that OT is the building block of multi-party secure
computations and more complicated \textquotedblleft post-cold-war
era\textquotedblright\ multi-party cryptographic protocols \cite{qi139}, and
quantum OT (QOT) can be obtained basing on quantum BC (QBC) \cite{qi75}. But
it is widely accepted that unconditionally secure QBC is impossible within
the quantum framework \cite{qi74}-\cite{qbc31}. This result, known as the
Mayers-Lo-Chau (MLC) no-go theorem, is considered as putting a serious
drawback on quantum cryptography. Obviously, it indicates that QOT built
upon QBC cannot be secure either. This stimulated the emergence of many
other no-go proofs on quantum two-party secure computations including QOT
\cite{qi149,qi500,qi797,qi499,qi677,qi725,qbc14,*qbc40,qbc61}.

Nevertheless, Kent showed that BC can be unconditionally secure under
relativistic settings \cite{qi44,qi582,qbc24,qbc51}. Recently, these
relativistic BC protocols were implemented experimentally \cite{qbc82,qbc83}%
. Also, 
there were many proposals on \textquotedblleft practical\textquotedblright\
QBC, which are secure if the participants are limited by some experimental
constraints, such as individual measurements or limited coherent
measurements, misaligned reference frames, unstability of particles,
Gaussian operations with non-Gaussian states, etc. (see the introduction of
Ref. \cite{HeJPA} for a detailed list).%

Therefore, it is natural to ask whether these BC protocols can lead to
secure QOT. That is, suppose that any setting or constraint required to
guarantee the security of the above BC protocols is satisfied, so that the
participants can use them as a secure \textquotedblleft black
box\textquotedblright\ without caring the internal details of these
protocols. Then we put no constraint (except those forbidden by fundamental
physics laws) on the participants' behaviors in other steps of the BC-based
QOT. Note that in this scenario, it may not be straight-forward to apply
some common methods adopted for proving the impossibility of QBC and related
no-go theorems, e.g., replacing any protocol with an ideal model which
contains quantum communications and unitary transformations only. This is
because now there is the secure \textquotedblleft black
box\textquotedblright\ QBC stands in the middle of the QOT process, so that
some cheating strategies may be interrupted at this stage. Thus it is
important to reexamine whether the no-go proofs of QOT still apply, and if
yes, how the cheating is performed.

In this paper, the answer is twofold. On one hand, we will provide a
cheating strategy in details, showing that some of the no-go proofs \cite%
{qi500,qi797,qi499,qi677,qi725,qbc14,*qbc40} remain valid even if QOT is
based on secure BC. On the other hand, we found that some other no-go proofs
\cite{qi149,qbc61} no longer work in such a QOT protocol, revealing that
these proofs are not sufficiently general.

\section{Definitions}

BC is a cryptographic task between two remote parties Charlie and Diana
(generally called Alice and Bob in literature. But to avoid confusing with
the roles in OT, here we name them differently). It generally includes two
phases. In the commit phase, Charlie decides the value of the bit $x$ ($x=0$
or $1$) which he wants to commit, and sends Diana a piece of evidence.
Later, in the unveil phase, Charlie announces the value of $x$, and Diana
checks it with the evidence. An unconditionally secure BC protocol needs to
be both binding (i.e., Charlie cannot change the value of $x$ after the
commit phase) and concealing (Diana cannot know $x$ before the unveil phase)
without relying on any computational assumption.

In the quantum case, Charlie's input can be more complicated. Besides the
two classical values $0$ and $1$, he can commit a quantum superposition or
mixture of the states corresponding to $x=0$\ and $x=1$, so that $x$ can be
unveiled as either $0$ or $1$\ with the probabilities $p_{0}$\ and $p_{1}$,
respectively. More specifically, suppose that a QBC protocol requires
Charlie to send Diana a quantum system $\Psi $ as the evidence in the commit
phase, whose state is expected to be $\left\vert \psi _{0}\right\rangle
_{\Psi }$ (if $x=0$) or $\left\vert \psi _{1}\right\rangle _{\Psi }$ (if $%
x=1 $). Then Charlie can introduce another system $C$, and prepare $C\otimes
\Psi $\ in the entangled state

\begin{equation}
\left\vert C\otimes \Psi \right\rangle =p_{0}^{1/2}\left\vert
c_{0}\right\rangle _{C}\otimes \left\vert \psi _{0}\right\rangle _{\Psi
}+p_{1}^{1/2}\left\vert c_{1}\right\rangle _{C}\otimes \left\vert \psi
_{1}\right\rangle _{\Psi },  \label{eqcheating}
\end{equation}%
where $\left\vert c_{0}\right\rangle _{C}$\ and $\left\vert
c_{1}\right\rangle _{C}$\ are orthogonal. He sends $\Psi $ to Diana and
keeps $C$ to himself. When it is time to unveil, Charlie measures $C$ in the
basis $\{\left\vert c_{0}\right\rangle _{C},\left\vert c_{1}\right\rangle
_{C}\}$, and unveils the committed $x$ as $0$ (or $1$) if the result is $%
\left\vert c_{0}\right\rangle _{C}$ (or $\left\vert c_{1}\right\rangle _{C}$%
). With this strategy, his commitment was kept at the quantum level until
the unveil phase, instead of taking a fixed classical value.

According to Kent \cite{qi581}, this \textquotedblleft is not considered a
security failure of a quantum BC protocol \textit{per se}\textquotedblright
. As long as a BC protocol can force Charlie to commit to a probability
distribution $(p_{0},p_{1})$\ which cannot be changed after the commit
phase, and $(p_{0}+p_{1})-1$\ can be made arbitrarily close to $0$ by
increasing some security parameters of the protocol, then it is still
considered as unconditionally secure. On the other hand, if a protocol can
further force Charlie to commit to a particular classical $x$, i.e., besides
$p_{0}+p_{1}\rightarrow 1$, both $p_{0}$ and $p_{1}$ can only take the
values $0$ or $1$ instead of any value in between, then it is called a bit
commitment with a certificate of classicality (BCCC). All the above
mentioned BC protocols \cite{qi44,qi582,qbc24,qbc51,HeJPA} are not BCCC, and
unconditionally secure BCCC seems impossible \cite{qi581}. Therefore, in the
following when speaking of secure BC, we refer to the non-BCCC ones only,
except where noted.

OT is also a two-party cryptography. There are two major types of OT in
literature. Using Cr\'{e}peau's description \cite{qi140}, they are defined
as follows.

\bigskip

\textit{Definition A: All-or-nothing OT (AoN OT)}

(A-i) Alice knows one bit $b$.

(A-ii) Bob gets bit $b$ from Alice with the probability $1/2$.

(A-iii) Bob knows whether he got $b$ or not.

(A-iv) Alice does not know whether Bob got $b$ or not.

\bigskip

\textit{Definition B: One-out-of-two OT (1-2 OT)}

(B-i) Alice knows two bits $b_{0}$ and $b_{1}$.

(B-ii) Bob gets bit $b_{j}$ and not $b_{\bar{j}}$ with $Pr(j=0)=Pr(j=1)=1/2$.

(B-iii) Bob knows which of $b_{0}$ or $b_{1}$ he got.

(B-iv) Alice does not know which $b_{j}$ Bob got.

\bigskip

We will study BC-based AoN OT first, and come back to 1-2 OT later.

\section{Insecurity}

According to Yao \cite{qi75}, AoN QOT can be built upon BC as follows.

\bigskip

\textit{The BC-based AoN QOT protocol:}

(I) Let $\left\vert 0,0\right\rangle $\ and $\left\vert 0,1\right\rangle $\
be two orthogonal states of a qubit, and define $\left\vert 1,0\right\rangle
\equiv (\left\vert 0,0\right\rangle +\left\vert 0,1\right\rangle )/\sqrt{2}$%
, $\left\vert 1,1\right\rangle \equiv (\left\vert 0,0\right\rangle
-\left\vert 0,1\right\rangle )/\sqrt{2}$. That is, the state of a qubit is
denoted as $\left\vert a_{i},g_{i}\right\rangle $, where $a_{i}$\ represents
the basis and $g_{i}$\ distinguishes the two states in the same basis. For $%
i=1,...,n$, Alice randomly picks $a_{i},g_{i}\in \{0,1\}$\ and sends Bob a
qubit $\phi _{i}$ in the state $\left\vert a_{i},g_{i}\right\rangle $.

(II) For $i=1,...,n$, Bob randomly picks a basis $b_{i}\in \{0,1\}$ to
measure $\phi_{i}$ and records the result as $\left\vert
b_{i},h_{i}\right\rangle $. Then he commits $(b_{i},h_{i})$ to Alice using
the BC protocol.

(III) Alice randomly picks a subset $R\subseteq \{1,...,n\}$\ and tests
Bob's commitment at positions in $R$. If any $i\in R$ reveals $a_{i}=b_{i}$
and $g_{i}\neq h_{i}$, then Alice stops the protocol; otherwise, the test
result is \textit{accepted}.

(IV) Alice announces the bases $a_{i}$\ ($i=1,...,n$). Let $T_{0}$ be the
set of all $1\leq i\leq n$ with $a_{i}=b_{i}$, and $T_{1}$ be the set of all
$1\leq i\leq n$ with $a_{i}\neq b_{i}$. Bob chooses $I_{0}\subseteq T_{0}-R$%
, $I_{1}\subseteq T_{1}-R$ with $\left\vert I_{0}\right\vert =\left\vert
I_{1}\right\vert =0.24n$, and sets $\{J_{0},J_{1}\}=\{I_{0},I_{1}\}$ or $%
\{J_{0},J_{1}\}=\{I_{1},I_{0}\}$ at random, then sends $\{J_{0},J_{1}\}$ to
Alice.

(V) Alice picks a random $s\in \{0,1\}$, and sends $s$, $\beta
_{s}=b\bigoplus\limits_{i\in J_{s}}g_{i}$ to Bob. Bob computes $b=\beta
_{s}\bigoplus\limits_{i\in J_{s}}h_{i}$ if $J_{s}=I_{0}$; otherwise does
nothing.

\bigskip

Now suppose that the BC protocol used in this QOT is secure. That is, no
matter we use relativistic BC \cite{qi44,qi582,qbc24,qbc51}, or
\textquotedblleft practical\textquotedblright\ QBC protocols listed in the
introduction of Ref. \cite{HeJPA}, we assume that all the security
requirements (e.g., relativistic settings or experimental limitations) are
already met, so that Bob does not have unlimited computational power to
cheat within the BC stage. In this case, the validity of the no-go proofs of
QOT \cite{qi149,qi500,qi797,qi499,qi677,qi725,qbc14,*qbc40,qbc61} cannot be
taken for granted, because all these proofs were derived without implying
any limitation on the computational power of the cheater.

Intriguingly, the conclusions of some of the no-go proofs \cite%
{qi500,qi797,qi499,qi677,qi725,qbc14,*qbc40} remain valid, that
unconditionally secure QOT is still impossible in this case. The key reason
is that secure BC, being not a BCCC, cannot avoid the participant keeping
the commitment at the quantum level instead of taking a fixed classical
value. Kent \cite{qi44} briefly mentioned that it will allow more general
coherent quantum attacks to be used on schemes of which BC is a subprotocol,
but no details of the cheating strategy was given. Here we will elaborate
how Bob can make use of this feature to break the BC-based QOT protocol.

For each $\phi _{i}$ ($i=1,...,n$), a dishonest Bob does not pick a
classical $b_{i}$ and measure it in step (II). Instead, he introduces two
ancillary qubit systems $B_{i}$ and $H_{i}$ as the registers for the bits $%
b_{i}$ and $h_{i}$, and prepares their initial states as $\left\vert
B_{i}\right\rangle =(\left\vert 0\right\rangle _{B}+\left\vert
1\right\rangle _{B})/\sqrt{2}$ and $\left\vert H_{i}\right\rangle
=\left\vert 0\right\rangle _{H}$,\ respectively. Here $\left\vert
0\right\rangle $ and $\left\vert 1\right\rangle $ are orthogonal. Then he
applies the unitary transformation%
\begin{eqnarray}
U_{1} &\equiv &\left\vert 0\right\rangle _{B}\left\langle 0\right\vert
\otimes \left\vert 0,0\right\rangle _{\phi }\left\langle 0,0\right\vert
\otimes I_{H}  \nonumber \\
&&+\left\vert 0\right\rangle _{B}\left\langle 0\right\vert \otimes
\left\vert 0,1\right\rangle _{\phi }\left\langle 0,1\right\vert \otimes
\sigma _{H}^{(x)}  \nonumber \\
&&+\left\vert 1\right\rangle _{B}\left\langle 1\right\vert \otimes
\left\vert 1,0\right\rangle _{\phi }\left\langle 1,0\right\vert \otimes I_{H}
\nonumber \\
&&+\left\vert 1\right\rangle _{B}\left\langle 1\right\vert \otimes
\left\vert 1,1\right\rangle _{\phi }\left\langle 1,1\right\vert \otimes
\sigma _{H}^{(x)}
\end{eqnarray}%
on the system $B_{i}\otimes \phi _{i}\otimes H_{i}$. Here $I_{H}$\ and $%
\sigma _{H}^{(x)}$\ are the identity operator and Pauli matrix of system $%
H_{i}$ that satisfy $I_{H}\left\vert 0\right\rangle _{H}=\left\vert
0\right\rangle _{H}$\ and $\sigma _{H}^{(x)}\left\vert 0\right\rangle
_{H}=\left\vert 1\right\rangle _{H}$,\ respectively. The effect of $U_{1}$
is like running a quantum computer program that if $\left\vert
B_{i}\right\rangle =\left\vert 0\right\rangle _{B}$ ($\left\vert
B_{i}\right\rangle =\left\vert 1\right\rangle _{B}$) then measures qubit $%
\phi _{i}$ in the basis $b_{i}=0$\ ($b_{i}=1$), and stores the result $h_{i}$
in system $H_{i}$. It differs from a classical program with the same
function as no destructive measurement is really performed, since $U_{1}$ is
not a projective operator. Consequently, the bits $b_{i}$ and $h_{i}$ are
kept at the quantum level instead of being collapsed to classical values.

Bob then commits $(b_{i},h_{i})$ to Alice at the quantum level. This can
always be done in a BC protocol which does not satisfy the definition of
BCCC. For example, to commit $b_{i}$, Bob further introduces two ancillary
systems $E$ and $\Psi $\ and prepares the initial state as

\begin{equation}
\left\vert E\otimes \Psi \right\rangle _{0}=\left\vert e_{0}\right\rangle
_{E}\otimes \left\vert \psi _{0}\right\rangle _{\Psi }.  \label{BC1}
\end{equation}%
Let $U_{E\otimes \Psi }$ be a unitary transformation on $E\otimes \Psi $\
satisfying $U_{E\otimes \Psi }\left\vert e_{0}\right\rangle _{E}\otimes
\left\vert \psi _{0}\right\rangle _{\Psi }=\left\vert e_{1}\right\rangle
_{E}\otimes \left\vert \psi _{1}\right\rangle _{\Psi }$. Here $\left\vert
\psi _{0}\right\rangle _{\Psi }$, $\left\vert \psi _{1}\right\rangle _{\Psi
} $\ have the same meanings as these in Eq. (\ref{eqcheating}), and $%
\left\vert e_{0}\right\rangle _{E}$, $\left\vert e_{1}\right\rangle _{E}$
are orthogonal. Bob applies the unitary transformation%
\begin{equation}
U_{2}\equiv \left\vert 0\right\rangle _{B}\left\langle 0\right\vert \otimes
I_{E\otimes \Psi }+\left\vert 1\right\rangle _{B}\left\langle 1\right\vert
\otimes U_{E\otimes \Psi }  \label{BC2}
\end{equation}%
on system $B_{i}\otimes E\otimes \Psi $, where $I_{E\otimes \Psi }$\ is the
identity operator of system $E\otimes \Psi $. As a result, we can see that
the final state of $B_{i}\otimes \phi _{i}\otimes H_{i}\otimes E\otimes \Psi
$\ will be very similar to Eq. (\ref{eqcheating}) if we view $B_{i}\otimes
\phi _{i}\otimes H_{i}\otimes E$\ as system $C$. Then Bob can follow the
process after Eq. (\ref{eqcheating}) (note that now Bob plays the role of
Charlie) to complete the commitment of $b_{i}$ without collapsing it to a
classical value. He can do the same to $h_{i}$.

Back to step (III) of the QOT protocol. Whenever $(b_{i},h_{i})$ ($i\in R$)
are picked to test the commitment, Bob simply unveils them honestly. Since
these $(b_{i},h_{i})$ will no longer be useful in the remaining steps of the
protocol, it does not hurt Bob's cheating. Note that the rest $(b_{i},h_{i})$
($i\notin R$) are still kept at the quantum level. After Alice announced all
bases $a_{i}$\ ($i=1,...,n$) in step (IV), Bob introduces a single global
control qubit $S^{\prime }$ for all $i$, initialized in the state $%
\left\vert s^{\prime }\right\rangle =(\left\vert 0\right\rangle _{S^{\prime
}}+\left\vert 1\right\rangle _{S^{\prime }})/\sqrt{2}$, and yet another
ancillary system $\Gamma _{i}$ for each $i\in T_{0}\cup T_{1}-R$ initialized
in the state $\left\vert \Gamma _{i}\right\rangle =\left\vert 0\right\rangle
_{\Gamma }$. Then he applies the unitary transformation%
\begin{eqnarray}
U_{3} &\equiv &\left\vert 0\right\rangle _{S^{\prime }}\left\langle
0\right\vert \otimes \left\vert a_{i}\right\rangle _{B}\left\langle
a_{i}\right\vert \otimes I_{\Gamma }  \nonumber \\
&&+\left\vert 0\right\rangle _{S^{\prime }}\left\langle 0\right\vert \otimes
\left\vert \lnot a_{i}\right\rangle _{B}\left\langle \lnot a_{i}\right\vert
\otimes \sigma _{\Gamma }^{(x)}  \nonumber \\
&&+\left\vert 1\right\rangle _{S^{\prime }}\left\langle 1\right\vert \otimes
\left\vert a_{i}\right\rangle _{B}\left\langle a_{i}\right\vert \otimes
\sigma _{\Gamma }^{(x)}  \nonumber \\
&&+\left\vert 1\right\rangle _{S^{\prime }}\left\langle 1\right\vert \otimes
\left\vert \lnot a_{i}\right\rangle _{B}\left\langle \lnot a_{i}\right\vert
\otimes I_{\Gamma }
\end{eqnarray}%
on the incremented system $S^{\prime }\otimes B_{i}\otimes \Gamma _{i}$.
Here $I_{\Gamma }$\ and $\sigma _{\Gamma }^{(x)}$\ are the identity operator
and Pauli matrix of system $\Gamma _{i}$ that satisfies $I_{\Gamma
}\left\vert 0\right\rangle _{\Gamma }=\left\vert 0\right\rangle _{\Gamma }$\
and $\sigma _{\Gamma }^{(x)}\left\vert 0\right\rangle _{\Gamma }=\left\vert
1\right\rangle _{\Gamma }$,\ respectively. The effect of $U_{3}$ is to
compare $a_{i}$ with $b_{i}$ and store the result $(a_{i}\neq b_{i})\oplus
s^{\prime }$\ in $\Gamma _{i}$. Bob then measures all $\Gamma _{i}$ ($i\in
T_{0}\cup T_{1}-R$) in the basis $\{\left\vert 0\right\rangle _{\Gamma
},\left\vert 1\right\rangle _{\Gamma }\}$, takes $T_{0}$ ($T_{1}$) as the
set of all $1\leq i\leq n$ with $\left\vert \Gamma _{i}\right\rangle
=\left\vert 0\right\rangle _{\Gamma }$ ($\left\vert \Gamma _{i}\right\rangle
=\left\vert 1\right\rangle _{\Gamma }$) instead of how they were defined in
step (IV), and always sets $J_{0}\subseteq T_{0}-R$, $J_{1}\subseteq T_{1}-R$
to finish the rest parts of the QOT protocol.

With this method, the relationship between $J_{0}$, $J_{1}$ and $I_{0}$, $%
I_{1}$\ are kept at the quantum level. Since $I_{0}$ ($I_{1}$) denotes the
set corresponding to $a_{i}=b_{i}$ ($a_{i}\neq b_{i}$). We can see that $%
U_{3}$ makes $J_{0}=I_{0}$, $J_{1}=I_{1}$\ when $s^{\prime }=0$, while $%
J_{0}=I_{1}$, $J_{1}=I_{0}$\ when $s^{\prime }=1$. As $S^{\prime }$ was
initialized as $\left\vert s^{\prime }\right\rangle =(\left\vert
0\right\rangle _{S^{\prime }}+\left\vert 1\right\rangle _{S^{\prime }})/%
\sqrt{2}$, the actual result of step (IV) can be described by the entangled
state%
\begin{eqnarray}
&&\left\vert S^{\prime }\otimes (\bigotimes\limits_{i}B_{i}\otimes \phi
_{i}\otimes H_{i}\otimes E_{i}^{\prime })\right\rangle  \nonumber \\
&\rightarrow &\left\vert \Phi _{b}\right\rangle =(\left\vert 0\right\rangle
_{S^{\prime }}\otimes \left\vert J_{0}=I_{0}\vee J_{1}=I_{1}\right\rangle
\nonumber \\
&&+\left\vert 1\right\rangle _{S^{\prime }}\otimes \left\vert
J_{0}=I_{1}\vee J_{1}=I_{0}\right\rangle )/\sqrt{2}.  \label{eqstepiv}
\end{eqnarray}%
Here $E_{i}^{\prime }$ stands for all the ancillary systems Bob introduced
in the process of committing $(b_{i},h_{i})$. $\left\vert J_{0}=I_{0}\vee
J_{1}=I_{1}\right\rangle $\ denotes the state of system $\bigotimes%
\limits_{i}B_{i}\otimes \phi _{i}\otimes H_{i}\otimes E_{i}^{\prime }$, in
which the subsystems $B_{i}$ and $H_{i}$\ contain the correct $b_{i}$ and $%
h_{i}$ corresponding to $J_{0}=I_{0}\vee J_{1}=I_{1}$. The meaning of $%
\left\vert J_{0}=I_{1}\vee J_{1}=I_{0}\right\rangle $\ is also similar.

After Alice announced $s$\ and $\beta _{s}$\ in step (V), the systems under
Bob's possession can be viewed as%
\begin{equation}
\left\vert \Phi _{b}\right\rangle =(\left\vert s\right\rangle _{S^{\prime
}}\otimes \left\vert J_{s}=I_{0}\right\rangle +\left\vert \lnot
s\right\rangle _{S^{\prime }}\otimes \left\vert fail\right\rangle )/\sqrt{2}.
\label{eqqot}
\end{equation}%
It means that if Bob measures system $S^{\prime }$ in the basis $%
\{\left\vert 0\right\rangle _{S^{\prime }},\left\vert 1\right\rangle
_{S^{\prime }}\}$\ and the result $\left\vert s^{\prime }\right\rangle
_{S^{\prime }}$ satisfies $s^{\prime }=s$, then he is able to measure the
rest systems and get all the correct $h_{i}$\ to decode the secret bit $b$
unambiguously; else if the result satisfies $s^{\prime }\neq s$, then he
knows that he fails to decode $b$. Now the most tricky part is, as the value
of $s^{\prime }$ was kept at the quantum level before system $S^{\prime }$
is measured, at this stage a dishonest Bob can choose not to measure $%
S^{\prime }$ in the basis $\{\left\vert 0\right\rangle _{S^{\prime
}},\left\vert 1\right\rangle _{S^{\prime }}\}$. Instead, by denoting $%
\left\vert b\right\rangle \equiv \left\vert s\right\rangle _{S^{\prime
}}\otimes \left\vert J_{s}=I_{0}\right\rangle $, and $\left\vert
?\right\rangle \equiv \left\vert \lnot s\right\rangle _{S^{\prime }}\otimes
\left\vert fail\right\rangle $,\ Eq. (\ref{eqqot}) can be treated as $%
\left\vert \Phi _{b}\right\rangle =(\left\vert b\right\rangle +\left\vert
?\right\rangle )/\sqrt{2}$ where $\left\vert b=0\right\rangle \equiv (%
\begin{array}{ccc}
1 & 0 & 0%
\end{array}%
)^{T}$, $\left\vert b=1\right\rangle \equiv (%
\begin{array}{ccc}
0 & 1 & 0%
\end{array}%
)^{T}$, and $\left\vert ?\right\rangle \equiv (%
\begin{array}{ccc}
0 & 0 & 1%
\end{array}%
)^{T}$ are mutually orthogonal. Then according to Eq. (33) of Ref. \cite%
{qi499}, Bob can distinguish them using the positive operator-valued measure
(POVM) $(E_{0},I-E_{0})$, where%
\begin{equation}
E_{0}=\frac{1}{6}\left[
\begin{array}{ccc}
2+\sqrt{3} & -1 & 1+\sqrt{3} \\
-1 & 2-\sqrt{3} & 1-\sqrt{3} \\
1+\sqrt{3} & 1-\sqrt{3} & 2%
\end{array}%
\right] .  \label{eqpovm}
\end{equation}%
This allows Bob's decoded $b$ to match Alice's actual input with reliability
$(1+\sqrt{3}/2)/2$ \cite{qi499}. On the contrary, when Bob executes the QOT
protocol honestly, in $1/2$ of the cases he can decode $b$ with reliability $%
100\%$; in the rest $1/2$ cases he fails to decode $b$, he can guess the
value randomly, which results in a reliability of $50\%$. Thus the average
reliability in the honest case is $100\%/2+50\%/2=75\%<(1+\sqrt{3}/2)/2$.
Note that in the above dishonest strategy, in any case Bob can never decode $%
b$ with reliability $100\%$. Therefore it is debatable whether it can be
considered as a successful cheating, as the strategy does not even
accomplish what an honest Bob can do. That is why it is called \textit{honest%
}-but-curious adversary \cite{qi677,qi725}, i.e., in some sense it may still
be regarded as honest behavior instead of full cheating. Nevertheless, it
provides Bob with the freedom to choose between accomplishing the original
goal of QOT or achieving a higher average reliability, which could leave
rooms for potential problems when building even more complicated
cryptographic protocols upon such a BC-based QOT.

The above cheating strategy is basically the same we proposed in section 5
of Ref. \cite{HeJPA}, which was applied to show why the specific QBC
protocol in the same reference cannot lead to secure QOT. But here we can
see that its power is not limited to the QBC protocol in Ref. \cite{HeJPA}.
Especially, Bob's steps related with Eqs. (\ref{BC1}) and (\ref{BC2}) will
always be valid as long as the BC protocol used in QOT is not a BCCC, as
they do not involve the details of the BC process. Thus we reach a much
general result, that any BC (except BCCC) cannot lead to unconditionally
secure AoN QOT using Yao's method \cite{qi75}. It covers not only
unconditionally secure QBC (regardless whether it exists or not), but also
relativistic BC (both classical \cite{qi44,qi582} and quantum ones \cite%
{qbc24,qbc51}) and practically secure QBC (e.g., those listed in the
introduction of Ref. \cite{HeJPA}), even if all the requirements for them to
be secure are already met. In this sense, QOT is more difficult than QBC, in
contrast to the classical relationship that OT and BC are quivalent.

This result shows that the original security proof of BC-based QOT \cite%
{qi75} is not general. The proof claimed that as long as the BC protocol is
unconditionally secure, then the QOT protocol built upon it will be
unconditionally secure too. But now we can see that it may still be valid
for BCCC-based QOT, but fails to cover all unconditionally secure BC.

Now consider 1-2 OT. It can be built upon BC in much the same way as the
above BC-based AoN QOT protocol, except that step (V) should be modified
into:

\bigskip

(V') Alice sends $\beta _{0}=b_{0}\bigoplus\limits_{i\in J_{0}}g_{i}$ and $%
\beta _{1}=b_{1}\bigoplus\limits_{i\in J_{1}}g_{i}$\ to Bob. Bob computes $%
b_{0}=\beta _{0}\bigoplus\limits_{i\in J_{0}}h_{i}$ if $J_{0}=I_{0}$, or $%
b_{1}=\beta _{1}\bigoplus\limits_{i\in J_{1}}h_{i}$ if $J_{1}=I_{0}$.

Bob can also apply the above cheating strategy, so that the result of step
(IV) is still described by Eq. (\ref{eqstepiv}). After Alice announced $%
\beta _{0}$ and $\beta _{1}$\ in step (V'), if Bob wants to decode $b_{0}$,
he can treat the right-hand side of Eq. (\ref{eqstepiv}) as%
\begin{equation}
\left\vert \Phi _{b}\right\rangle =(\left\vert 0\right\rangle _{S^{\prime
}}\otimes \left\vert J_{0}=I_{0}\right\rangle +\left\vert 1\right\rangle
_{S^{\prime }}\otimes \left\vert fail\right\rangle )/\sqrt{2},
\label{eqqot1}
\end{equation}%
else if he wants to decode $b_{1}$, he can treat it as%
\begin{equation}
\left\vert \Phi _{b}\right\rangle =(\left\vert 0\right\rangle _{S^{\prime
}}\otimes \left\vert fail\right\rangle +\left\vert 1\right\rangle
_{S^{\prime }}\otimes \left\vert J_{1}=I_{0}\right\rangle )/\sqrt{2}.
\label{eqqot2}
\end{equation}%
Comparing these two equations with Eq. (\ref{eqqot}), we can see that they
both have the form $\left\vert \Phi _{b}\right\rangle =(\left\vert
b\right\rangle +\left\vert ?\right\rangle )/\sqrt{2}$. Thus Bob can still
apply the POVM described by Eq. (\ref{eqpovm}) to decode the bit he wants.
Consequently, he can decode one of $b_{0}$ and $b_{1}$ at his choice with
reliability $(1+\sqrt{3}/2)/2$. Again, despite that the value is higher than
the average reliability of the honest behavior, in the current case Bob can
never decode the bit with reliability $100\%$. Thus it still belongs to the
honest-but-curious adversaries. Also, it is important to note that the POVM $%
(E_{0},I-E_{0})$ is a two-value measurement that can obtain one bit of
information only, and the POVMs corresponding to Eq. (\ref{eqqot1}) and Eq. (%
\ref{eqqot2}) are not the same. Therefore Bob can pick only one of them to
increase the average reliability of one of $b_{0}$ and $b_{1}$, instead of
decoding both bits simultaneously.

From the above cheating strategies, we can see that Bob's key idea is to
keep introducing quantum entanglement to the system, which enables him to
keep more and more data at the quantum level, so that he can have the
freedom on choosing different measurements at a later time. This gives yet
another example showing the power of entanglement in quantum cryptography.

\section{Security}

The above honest-but-curious adversaries indicate that the BC-based QOT
protocol is not unconditionally secure, which is in agreement with the
conclusion of the no-go proofs of QOT \cite%
{qi500,qi797,qi499,qi677,qi725,qbc14,*qbc40}. Nevertheless, we will show
below that this protocol is secure against the cheating strategy in other
no-go proofs \cite{qi149,qbc61}.

In Lo's no-go proof \cite{qi149}, the following definition of 1-2 OT was
proposed.

\bigskip

\textit{Definition C: Lo's 1-2 OT}

(C-i) Alice inputs $i$, which is a pair of messages $(m_{0},m_{1})$.

(C-ii) Bob inputs $j=0$ or $1$.

(C-iii) At the end of the protocol, Bob learns about the message $m_{j}$,
but not the other message $m_{\bar{j}}$, i.e., the protocol is an ideal
one-sided two-party secure computation $f(m_{0},m_{1},j=0)=m_{0}$\ and $%
f(m_{0},m_{1},j=1)=m_{1}$.

(C-iv) Alice does not know which $m_{j}$ Bob got.

\bigskip

It was introduced as a special case of the ideal one-sided two-party quantum
secure computations, defined in Lo's proof as follows.

\bigskip

\textit{Definition D: ideal one-sided two-party secure computation}

Suppose Alice has a private (i.e. secret) input $i\in \{1,2,...,n\}$ and Bob
has a private input $j\in \{1,2,...,m\}$. Alice helps Bob to compute a
prescribed function $f(i,j)\in \{1,2,...,p\}$ in such a way that, at the end
of the protocol:

(a) Bob learns $f(i,j)$ unambiguously;

(b) Alice learns nothing [about $j$\ or $f(i,j)$];

(c) Bob knows nothing about $i$ more than what logically follows from the
values of $j$ and $f(i,j)$.

\bigskip

Lo's proof \cite{qi149} showed that any protocol satisfying Definition D is
insecure, because Bob can always obtain all $f(i,j)$ ($j\in \{1,2,...,m\}$).
As a corollary, secure 1-2 OT satisfying Definition C is impossible, as Bob
can always learn both $m_{0}$ and $m_{1}$.

This result is surprising. As shown in the previous section, other no-go
proofs \cite{qi500,qi797,qi499,qi677,qi725,qbc14,*qbc40} claimed that QOT is
insecure, merely because Bob can increase the average reliability of the
decoded value of one of $m_{0}$ and $m_{1}$. It is never indicated in Refs.
\cite{qi500,qi797,qi499,qi677,qi725,qbc14,*qbc40} that he can decode both of
them simultaneously. Thus the cheating strategy in Lo's proof \cite{qi149}
seems more powerful.

However, it will be shown below that Lo's proof is not sufficiently general
to cover all kinds of QOT. We must notice that Definition C is not
rigorously equivalent to Definition B. An important feature of Definition C
is that all Alice's (Bob's) input to the entire protocol is merely $%
i=\{m_{0},m_{1}\}$ ($j=\{0,1\}$). Furthermore, as can be seen from (C-i) and
(C-iii), the inputs $i$ and $j$ are independent of each other. But in
general, seldom any protocol satisfies these requirement. That is, let us
denote all Alice's (Bob's) input to a protocol as $I$ ($J$). In Definition C
there is $I=i$, $J=j$, and $I$, $J$ are independent. But most existing
quantum cryptographic protocols generally have $I\supset i$, $J\supset j$,
and $I$, $J$ are dependent of each other.

For example, in the well-known Bennett-Brassard 1984 (BB84) QKD protocol
\cite{qi365}, though the aim of Alice and Bob is to share a secret key $k$,
the protocol cannot be modeled as a simple box to which Alice inputs $k$,
then Bob gets the output $k$. Instead, more inputs of both participants have
to be involved. Alice should first input some quantum states (denoted as
input $i_{1}$), and Bob inputs and announces his measurement bases (input $%
j_{1}$). Then Alice tells Bob which bases are correct (input $i_{2}$),
followed by a security check in which Bob reveals some measurement results
(input $j_{2}$), and Alice verifies whether these results are correct or not
(input $i_{3}$). Alice also reveals some results for Bob to verify \ldots\
Finally they obtain $k$ from the remaining unannounced measurement results.
Obviously Alice cannot determine $i_{2}$ without knowing $j_{1}$, Bob's $%
j_{2}$ will be affected by Alice's $i_{1}$, \ldots\ , the final key $k$ is
also affected by the $i$'s and $j$'s. Thus we see that in the BB84 protocol,
the inputs $I=\{i_{1},i_{2},\ldots \}$ and $J=\{j_{1},j_{2},\ldots \}$ are
dependent of each other. For an eavesdropper, even though parts of $I$ and $%
J $ are revealed, it is still insufficient to decode $k$.

This is also the case for OT. Alice and Bob generally need to send quantum
states, perform operations and exchanges lots of information throughout the
entire protocol. All these (e.g., Alice's $\{a_{i},g_{i}\}$, $R$, $\beta
_{0} $, $\beta _{1}$\ and Bob's $\{b_{i},h_{i}\} $, $\{J_{0},J_{1}\}$\ in
the protocol in section III) should be treated as parts of their inputs.
Consequently, there is $I\supset i$ and $J\supset j $. Definition B requires
that Alice has zero knowledge about $j$. But it does not necessarily imply
that she has zero knowledge about $J$. Therefore $I$ and $J$ can be
dependent of each other. Indeed, step (V') of the BC-based 1-2 QOT protocol
in section III clearly shows that $I$ includes not only the secret bits $b_{0}$
and $b_{1}$, but also depends on how Bob selects $J_{0}$ and $J_{1}$ in step
(IV). Meanwhile, Bob's announcing $J_{0}$ and $J_{1}$ does not necessarily
reveal his choice of $j$. Therefore, comparing with Definitions C and D, the
BC-based 1-2 QOT protocol cannot be viewed as an ideal\ function $%
f(i(m_{0},m_{1}),j)$, where $i$ and $j$ are merely the private inputs of
Alice and Bob, respectively. Instead, it has the form $f(I(m_{0},m_{1},J),J)$%
, where Alice' input $I$ will be varied according to Bob's input $J$, and
its value is not determined until Bob's input has been completed. That is,
BC-based 1-2 QOT does not satisfy Definition C.

With this feature, the cheating strategy in Lo's proof can be defeated, as
it was pointed out in Ref. \cite{2OT} which will be reviewed below.
According to Lo's strategy, Bob can cheat in 1-2 OT satisfying Definition C,
because he can change the value of $j$ from $j_{1}$ to $j_{2}$ by applying a
unitary transformation to his own quantum machine alone. This enables him to
learn $f(i(m_{0},m_{1}),j_{1})$\ and $f(i(m_{0},m_{1}),j_{2})$\
simultaneously without being found by Alice. However, in a protocol
described by the function $f(I(m_{0},m_{1},J),J)$, a value in the form $%
f(I(m_{0},m_{1},J_{(1)}),J_{(2)})$\ (with $J_{(k)}$\ denoting Bob's input
corresponding to $j_{k}$) will be meaningless. Without the help of Alice,
Bob cannot change $I$ from $I(m_{0},m_{1},J_{(1)})$\ to $%
I(m_{0},m_{1},J_{(2)})$. Hence he cannot learn $%
f(I(m_{0},m_{1},J_{(1)}),J_{(1)})$\ and $f(I(m_{0},m_{1},J_{(2)}),J_{(2)})$\
simultaneously by\ himself. Thus the BC-based 1-2 QOT protocol is immune to
this cheating.

Now we prove it in a more rigorous mathematical form, following the
procedure in the appendix of Ref. \cite{2OT}. According to the cheating
strategy in Lo's proof as shown in section III of Ref. \cite{qi149}, in any
protocol satisfying Definition D, Alice and Bob's actions on their quantum
machines can be summarized as an overall unitary transformation $U$ applied
to the initial state $\left\vert u\right\rangle _{in}\in H_{A}\otimes H_{B}$%
, i.e.
\begin{equation}
\left\vert u\right\rangle _{fin}=U\left\vert u\right\rangle _{in}.
\label{e1}
\end{equation}%
When both parties are honest, $\left\vert u^{h}\right\rangle
_{in}=\left\vert i\right\rangle _{A}\otimes \left\vert j\right\rangle _{B}$
and
\begin{equation}
\left\vert u^{h}\right\rangle _{fin}=\left\vert v_{ij}\right\rangle \equiv
U(\left\vert i\right\rangle _{A}\otimes \left\vert j\right\rangle _{B}).
\label{e2}
\end{equation}%
Thus the density matrix that Bob has at the end of protocol is
\begin{equation}
\rho ^{i,j}=Tr_{A}\left\vert v_{ij}\right\rangle \left\langle
v_{ij}\right\vert .  \label{e3}
\end{equation}

Bob can cheat in this protocol, because given $j_{1},j_{2}\in \{1,2,...,m\}$%
, there exists a unitary transformation $U^{j_{1},j_{2}}$ such that
\begin{equation}
U^{j_{1},j_{2}}\rho ^{i,j_{1}}(U^{j_{1},j_{2}})^{-1}=\rho ^{i,j_{2}}
\label{e4}
\end{equation}%
for all $i$. It means that Bob can change the value of $j$ from $j_{1}$ to $%
j_{2}$ by applying a unitary transformation independent of $i$\ to the state
of his quantum machine. This equation is derived as follows \cite{qi149}.

Alice may entangle the state of her quantum machine $A$ with her quantum
dice $D$ and prepares the initial state
\begin{equation}
\frac{1}{\sqrt{n}}\sum\limits_{i}\left\vert i\right\rangle _{D}\otimes
\left\vert i\right\rangle _{A}.  \label{e5}
\end{equation}%
She keeps $D$ for herself and uses the second register $A$ to execute the
protocol. Supposing that Bob's input is $j_{1}$, the initial state is
\begin{equation}
\left\vert u^{\prime }\right\rangle _{in}=\frac{1}{\sqrt{n}}%
\sum\limits_{i}\left\vert i\right\rangle _{D}\otimes \left\vert
i\right\rangle _{A}\otimes \left\vert j_{1}\right\rangle _{B}.  \label{e6}
\end{equation}%
At the end of the protocol, it follows from Eqs. (\ref{e1}) and (\ref{e6})
that the total wave function of the combined system $D$, $A$, and $B$ is
\begin{equation}
\left\vert v_{j_{1}}\right\rangle _{in}=\frac{1}{\sqrt{n}}%
\sum\limits_{i}\left\vert i\right\rangle _{D}\otimes U(\left\vert
i\right\rangle _{A}\otimes \left\vert j_{1}\right\rangle _{B}).  \label{e7}
\end{equation}%
Similarly, if Bob's input is $j_{2}$, the total wave function at the end
will be
\begin{equation}
\left\vert v_{j_{2}}\right\rangle _{in}=\frac{1}{\sqrt{n}}%
\sum\limits_{i}\left\vert i\right\rangle _{D}\otimes U(\left\vert
i\right\rangle _{A}\otimes \left\vert j_{2}\right\rangle _{B}).  \label{e8}
\end{equation}%
Due to the requirement (b) in Definition D, the reduced density matrices in
Alice's hand for the two cases $j=j_{1}$ and $j=j_{2}$ must be the same,
i.e.
\begin{equation}
\rho _{j_{1}}^{Alice}=Tr_{B}\left\vert v_{j_{1}}\right\rangle \left\langle
v_{j_{1}}\right\vert =Tr_{B}\left\vert v_{j_{2}}\right\rangle \left\langle
v_{j_{2}}\right\vert =\rho _{j_{2}}^{Alice}.  \label{e9}
\end{equation}%
Equivalently, $\left\vert v_{j_{1}}\right\rangle $ and $\left\vert
v_{j_{2}}\right\rangle $\ have the same Schmidt decomposition
\begin{equation}
\left\vert v_{j_{1}}\right\rangle =\sum\limits_{k}a_{k}\left\vert \alpha
_{k}\right\rangle _{AD}\otimes \left\vert \beta _{k}\right\rangle _{B}
\label{e10}
\end{equation}%
and
\begin{equation}
\left\vert v_{j_{2}}\right\rangle =\sum\limits_{k}a_{k}\left\vert \alpha
_{k}\right\rangle _{AD}\otimes \left\vert \beta _{k}^{\prime }\right\rangle
_{B}.  \label{e11}
\end{equation}%
Now consider the unitary transformation $U^{j_{1},j_{2}}$ that rotates $%
\left\vert \beta _{k}\right\rangle _{B}$\ to $\left\vert \beta _{k}^{\prime
}\right\rangle _{B}$. Notice that it acts on $H_{B}$ alone and yet, as can
be seen from Eqs. (\ref{e10}) and (\ref{e11}), it rotates $\left\vert
v_{j_{1}}\right\rangle $ to $\left\vert v_{j_{2}}\right\rangle $,\ i.e.
\begin{equation}
\left\vert v_{j_{2}}\right\rangle =U^{j_{1},j_{2}}\left\vert
v_{j_{1}}\right\rangle .  \label{e12}
\end{equation}%
Since
\begin{equation}
_{D}\left\langle i\right. \left\vert v_{j}\right\rangle =\frac{1}{\sqrt{n}}%
\left\vert v_{ij}\right\rangle  \label{e13}
\end{equation}%
[see Eqs. (\ref{e2}), (\ref{e7}), and (\ref{e8})], by multiplying Eq. (\ref%
{e12}) by $_{D}\left\langle i\right\vert $\ on the left, one finds that
\begin{equation}
\left\vert v_{ij_{2}}\right\rangle =U^{j_{1},j_{2}}\left\vert
v_{ij_{1}}\right\rangle .  \label{e14}
\end{equation}%
Taking the trace of $\left\vert v_{ij_{2}}\right\rangle \left\langle
v_{ij_{2}}\right\vert $\ over $H_{A}$ and using Eq. (\ref{e14}), Eq. (\ref%
{e4}) can be obtained.

Eqs. (\ref{e1}) - (\ref{e14}) are exactly those presented in Lo's proof \cite%
{qi149}. We now consider the BC-based 1-2 QOT protocol. Since it has the
feature that Alice's input $I$\ is dependent of Bob's input $J$, in the
above proof, all $i$ in the equations should be replaced by $I(J)$ from the
very beginning. Consequently, Eq. (\ref{e13}) becomes
\begin{equation}
_{D}\left\langle I(J)\right\vert \left. v_{J}\right\rangle =\frac{1}{\sqrt{n}%
}\left\vert v_{I(J)J}\right\rangle .  \label{e15}
\end{equation}%
In this case, multiplying Eq. (\ref{e12}) by $_{D}\left\langle
I_{(2)}\right\vert $\ ($I_{(2)}\equiv I(J_{(2)})$ for short) on the left
cannot give Eq. (\ref{e14}) any more. Instead, the result is
\begin{equation}
\left\vert v_{I_{(2)}J_{(2)}}\right\rangle
=U^{J_{(1)},J_{(2)}}U^{I_{(1)},I_{(2)}}\left\vert
v_{I_{(1)}J_{(1)}}\right\rangle ,  \label{e16}
\end{equation}%
where $U^{I_{(1)},I_{(2)}}\equiv _{D}\left\vert I_{(2)}\right\rangle
\left\langle I_{(1)}\right\vert _{D}$. Then Eq. (\ref{e4}) is replaced by
\begin{equation}
U^{J_{(1)},J_{(2)}}U^{I_{(1)},I_{(2)}}\rho
^{I_{(1)},J_{(1)}}(U^{J_{(1)},J_{(2)}}U^{I_{(1)},I_{(2)}})^{-1}=\rho
^{I_{(2)},J_{(2)}}.  \label{e17}
\end{equation}%
Note that $U^{I_{(1)},I_{(2)}}$ is the unitary operation on Alice's side.
This implies that without Alice's help, Bob cannot change the density matrix
he has from $\rho ^{I_{(1)},J_{(1)}}$\ to $\rho ^{I_{(2)},J_{(2)}}$. That is
why Bob's cheating strategy fails.

In brief, Lo's no-go proof on ideal one-sided two-party secure computations
\cite{qi149} cannot cover the above BC-based 1-2 QOT, \ because the proof
studied merely the protocols in which the inputs of the participants are
independent. As we mentioned, even the BB84 protocol does not satisfy this
requirement, while it can still be used as a black box to build more
sophisticated protocols, e.g., quantum secret sharing. Thus we see that
black box protocols do not necessarily require independent inputs of the
participants. The model used in Lo's proof is too ideal, so that many useful
protocols in quantum cryptography are not covered.

Similarly, a recent no-go proof on two-sided two-party secure computations
\cite{qbc61} is also based on a model of protocols with independent inputs.
Moreover, the proof contains a logical loophole on the use of the security
definition \cite{CommentQBC61}. Therefore its conclusion is not sufficiently
general either.

\section{Summary and discussions}

We elaborated how Bob can make use of quantum entanglement to break the
above BC-based QOT and achieve a higher average reliability of the decoded
value of the transferred bit, even under certain practical settings in which
the no-go proofs for secure QBC become invalid. Meanwhile, we also showed
that BC-based QOT, though not unconditionally secure, can defeat certain
kinds of cheating which attempt to decode all transferred bits
simultaneously with reliability $100\%$. Thus it is still valuable for
building some \textquotedblleft post-cold-war era\textquotedblright\ quantum
cryptographies.

This insecurity proof is valid as long as the secure BC used in the QOT
protocol is not BCCC. It covers relativistic BC \cite{qi44,qi582,qbc24,qbc51}%
, as well as many practically secure QBC \cite{qi295,qi669,qbc43},
conditionally secure QBC \cite{qi63}, computationally secure QBC \cite%
{qbc42,qbc21,qbc34}, cheat-sensitive QBC \cite%
{qi150,qbc50,qbc78,qi197,qbc52,qbc89}, and some other types of protocols
\cite{HeJPA,HeQIP,HePRA,HeProof}. Nevertheless, when Bob is limited to
bounded or noisy quantum storages, secure QOT can be made possible in
practice with two approaches. On one hand, with this technological
constraint BCCC can be obtained \cite%
{qbc26,qi137,qi243,qbc41,qbc39,qi796,qbc59,qbc65,qbc98}. This is because Bob
can no longer keep system $C$\ in Eq. (\ref{eqcheating}) (which represents
the systems $B_{i}\otimes \phi _{i}\otimes H_{i}\otimes E$\ mentioned below
Eq. (\ref{BC2}) or $B_{i}\otimes \phi _{i}\otimes H_{i}\otimes E_{i}^{\prime
}$\ in Eq. (\ref{eqstepiv})) perfectly in the entangled form shown by these
equations once the protocol lasts too long or requires too much quantum
storages. Thus BC-based QOT protocol in section III can be secure at least
against the insecurity proof in this paper. On the other hand, bounded or
noisy storages can also force Bob to measure the quantum states he receives,
as long as the QOT protocol involves too much qubits or the time interval
between each step is sufficiently long. BC is no longer needed to convince
Alice that Bob has already completed the measurements. Then there can be QOT
protocols not based on BC, which are proven to be practically secure \cite%
{qi243,qbc41,qbc39,qi796,qi795,qbc6,qbc86}.

We should also note that, even without the assumption of such technological
limitations, our above insecurity proof does not mean that all QOT must not
be unconditionally secure in principle. This is because the existing method
\cite{qi75} is not necessarily the only way to build OT from BC. Further
more, there is no evidence indicating that OT has to be built upon BC.
Therefore, it is still worth questioning whether other kinds of
unconditionally secure OT exist, especially relativistic OT. \newline

The work was supported in part by 
the NSF of Guangdong province.

\end{document}